\documentclass[12pt]{article}

\begin{document}

\begin{center}

{\large {\bf A Class of Anisotropic Five-Dimensional Solutions for the Early Universe}}
\end{center}

\vskip 1.0 cm

\begin{center}
{ J. Ponce de Leon$^{1 \dagger}$ and P.S. Wesson$^{2 \ast}$}
\vskip 1.2 cm
{\it $^1$Department of Physics, 
University of Puerto Rico, P.O. Box 23343, San Juan, PR 00931, USA.\\}
\vskip 0.1cm
{\it $^2$Department of Physics and Astronomy, University of Waterloo, Waterloo, Ontario N2L 3G1, Canada.\\}

\vskip 0.6cm

{September   2008 (Revised version)}

\end{center}

\bigskip

\begin{abstract}

\bigskip

We solve the Ricci-flat equations of extended general relativity to obtain an interesting class of cosmological models. The solutions are analogous to the $4D$ ones of Bianchi type-I of Kasner type and have significant implications for astrophysics.

\end{abstract}

\bigskip



\bigskip

\bigskip

\bigskip

\bigskip

\bigskip

\bigskip

\bigskip

\bigskip

\bigskip

\bigskip

\bigskip

\bigskip

\bigskip

\bigskip

\bigskip

\bigskip

\bigskip

--------------------------------------------------------------

$^{\dagger}$E-mail: jpdel@ltp.upr.clu.edu or jpdel1@hotmail.com

$^{\ast}$E-mail: wesson@astro.uwaterloo.ca or pswastro@hotmail.com

Both authors belong to The S.T.M. Consortium, http://astro.uwaterloo.ca/$\sim$wesson

\newpage

\section{Introduction} 
   
Standard cosmology based on the general theory of relativity uses the Friedmann-Robertson-Walker (FRW) models, which are perfectly isotropic and homogeneous, or in short `uniform'. But many workers have suggested that the early universe may not have been exactly uniform, especially since it must have harbored the inhomogeneities which later evolved to give the galaxies and other structures we observe.

The simplest extension of the FRW models is the anisotropic ones of Bianchi type I, notably that of Kasner \cite{Kasner}. In such models, it is possible for the three spatial dimensions to undergo repeated ``bounces",  providing a description for the big-bang fireball if it was initially chaotic and evolved towards uniformity. (The use of a $4D$ Kasner-type metric in this manner 
leads to the so-called mixmaster universe as discussed in ref. $2$,  but we defer a discussion of the $5D$ analog to another place.) Many anisotropic models for the early universe have been much studied within the context of the usual four-dimensional theory of gravity due to Einstein. However, extensions of that theory to five and more dimensions are currently seen as providing the best route to unification of gravity with the interactions of particle physics \cite{particle physics}. The two versions of $5D$ relativity in vogue are membrane theory and space-time-matter (or induced matter) theory. Both employ a $5D$ Kaluza-Klein type of metric, where the extra dimension is not assumed to be compactified as in the original account and is indeed algebraically general \cite{algebraically}, \cite{general}. These two modern approaches to $5D$ relativity are actually equivalent in a mathematical sense, but are different as regards physical motivation \cite{physical motivation}. Membrane theory allows gravity to propagate in all $5$-dimensions, whereas the interactions of particles are confined to a singular $4D$ hypersurface, thereby accounting for their relatively greater strength \cite{strength}. Space-time-matter theory views the $5D$ world as pure geometry, where the matter we see in $4D$ is a consequence of the extra dimension in accordance with Campbell's embedding  theorem \cite{Campbell}, \cite{Seahra}. Both theories are in agreement with observations, and both have large literatures of which reviews are available \cite{review}, \cite{available}. Our present work is applicable to either approach, and is specific: we will derive an exact class of solutions to the $5D$ field equations which describe anisotropic cosmological models of Bianchi type-I that may be applicable to the early universe. 

Our solutions will prove to be generalizations of other $5D$ ones \cite{other}-\cite{ones} and reduce to the $4D$ Bianchi type-I models in the appropriate limits \cite{limits}. The solutions derived bellow generalize, most notably, the `standard' $5D$ cosmologies \cite{standard}. Those involve the fifth coordinate in a significant way, but have $4D$ subspaces with FRW-like properties. They were discovered using the method of separation of variables, which ensures that they reduce to the appropriate $4D$ solutions on hypersurfaces defined by fixing the fifth coordinate so as to recover spacetime. These solutions have one assignable parameter, which determines both the $3D$ expansion rate and the properties of the matter as defined by an effective energy-momentum tensor that describes various forms of matter, including radiation (early universe) and dust (late universe). The standard $5D$ solutions also include the analog of the de Sitter solution, which via the cosmological constant describes a vacuum-dominated phase (very early universe). However, we will not be much concerned with this phase in what follows, because it has recently been the subject of intense investigation \cite{intense}. Instead, we will concentrate on the early universe, and use the method of separation of variables to obtain exact solutions to the field equations in $5D$ that have physically acceptable properties of matter in $4D$.

For the field equations, we will follow the usual approach and employ the $5D$ Ricci tensor in the form $R_{AB} = 0$ $(A, B = 0, 1, 2, 3, 4)$. These  apparently empty solutions in $5D$ may contain matter in $4D$, because they include as a subset the Einstein field equations $G_{\alpha \beta} = 8 \pi T_{\alpha \beta}$ $(\alpha, \beta = 0, 1, 2, 3)$. This is a consequence of the aforementioned theorem of Campbell, which provides a local embedding via the formal rules of differential geometry that can be interpreted in a physical way to connect the Einstein tensor $(G_{\alpha \beta})$ and an effective energy-momentum-tensor  $(T_{\alpha \beta})$. This method for ensuring compliance with the field equations of general relativity has been widely applied. We will employ it briefly bellow, but most of our work will be concerned with the ``raw" solutions and their associated metrics. The latter will be defined via a $5D$ line element $dS^2 = \gamma_{AB} dx^{A}dx^{B}$, which contains the conventional $4D$ counterpart $ds^2 = g_{\alpha\beta}dx^{\alpha}dx^{\beta}$. The spacetime coordinates will be assigned as usual, with $x^{0} = t$ for time and $x^{1, 2, 3} = x, y, z$ for space. We will write $x^{4} = \psi$ for the extra coordinate, in order to avoid confusion with the spatial Cartesian label and to avoid the implication that $x^{4}$ is measured with respect to some special hypersurface. The gravitational constant $G$ and the speed of light $c$ will be set unity by an appropriate choice of units. Other aspects of our terminology are standard. Most of our results are algebraic in nature and derived in Section $2$, but we mention some astrophysical implications \cite{astrophysical}-\cite{implications} in Section $3$. For rushed readers, equation (\ref{Kasner-like metric in 5D, t and psi dependence}) defines our class of solutions.  

\section{Five-dimensional Bianchi Type-I Solutions}

In this section we wish to solve the field equations $R_{AB} = 0$ using the method of separation of variables in $x^{0} = t$ and $x^{4} = \psi$.

We start with a $5D$ line element where the components of the metric tensor are independent of the coordinates for ordinary $3D$ space $(x, y, z)$ and split via two functions $Q_{A}(\psi)$ and $S_{A}(t)$. Thus, we write

\begin{equation}
\label{general line element}
dS^2 = Q^2_{0}(\psi)S_{0}^2(t) dt^2 - Q^2_{1}(\psi)S^2_{1}(t)dx^2 - Q^2_{2}(\psi)S^2_{2}(t)dy^2 - Q^2_{3}(\psi)S^2_{3}(t)dz^2 + \epsilon Q^2_{4}(\psi)S^2_{4}(t)d\psi^2.
\end{equation}
Here $\epsilon = \pm 1$ allows for a spacelike or timelike extra dimension, both of which are physically admissible. (See, e.g.,  references \cite{general}, \cite{Seahra}, \cite{JPdeL}: in modern noncompactified $5D$ theory, the extra coordinate is different in nature from the spacetime coordinates, so there is no problem with closed timelike paths). 

Our starting metric (1) is, of course, somewhat special in terms of 5D algebra; but it will be seen to be adequate in terms of yielding the appropriate 4D physics.  Bearing in mind the opportunity to simplify (1), we  get rid of the gauge functions  $S_{0}(t)$ and $Q_{4}(\psi)$ by means of the  coordinate transformations

\begin{equation}
\label{transformation of coordinates}
S_{0}(t) dt \rightarrow d\bar{t}, \;\;\;\;\mbox{and}\;\;\;\;Q_{4}(\psi) d\psi \rightarrow d \bar{\psi}.
\end{equation}
Then relabeling the coordinates (dropping the overbars) the metric (\ref{general line element}) becomes

\begin{equation}
\label{line element}
dS^2 = Q^2_{0}({\psi})d{t}^2 - Q^2_{1}({\psi})S^2_{1}({t})dx^2 - Q^2_{2}({\psi})S^2_{2}({t})dy^2 - Q^2_{3}({\psi})S^2_{3}({t})dz^2 + \epsilon S^2_{4}({t})d{\psi^2}.
\end{equation}
This metric reduces to homogeneous and isotropic Bianchi type-I cosmological models with flat spatial sections on every hypersurface $\Sigma_{\psi}: \psi = \psi_{0} = $ constant. There are eight unknown functions in (\ref{line element}), which are to be determined from $R_{AB} = 0$.

The field equations have in fact only six nonzero components, which it is convenient to consider in the order $R_{00}, R_{44}, R_{04}, R_{11}, R_{22}, R_{33}$. Setting these to zero gives the following set of equations:

\begin{equation}
\label{R00}
\epsilon S_{4}^2\left[\frac{\ddot{S}_{1}}{S_{1}} + \frac{\ddot{S}_{2}}{S_{2}} + \frac{\ddot{S}_{3}}{S_{3}} + \frac{\ddot{S}_{4}}{S_{4}}\right] + Q_{0}^2\left[\frac{{{Q}}_{0}''}{Q_{0}} + \frac{{{Q}}_{0}'}{Q_{0}}\left(\frac{{Q}_{1}'}{Q_{1}} + \frac{{Q}_{2}'}{Q_{2}} + \frac{{Q}_{3}'}{Q_{3}}\right)\right] = 0,
\end{equation}

\bigskip

\begin{equation}
\label{R44}
\epsilon S_{4}^2\left[\frac{{\ddot{S}}_{4}}{S_{4}} + \frac{{\dot{S}}_{4}}{S_{4}}\left(\frac{\dot{S}_{1}}{S_{1}} + \frac{\dot{S}_{2}}{S_{2}} + \frac{\dot{S}_{3}}{S_{3}}\right)\right] + Q_{0}^2\left[\frac{{Q}''_{0}}{Q_{0}} + \frac{{Q}''_{1}}{Q_{1}} + \frac{{Q}''_{2}}{Q_{2}} + \frac{{Q}''_{3}}{Q_{3}}\right]  = 0,
\end{equation}

\bigskip

\begin{equation}
\label{R04}
- \frac{Q'_{0}}{Q_{0}}\left(\frac{{\dot{S}}_{1}}{S_{1}} + \frac{{\dot{S}}_{2}}{S_{2}} + \frac{{\dot{S}}_{3}}{S_{3}}\right) + \frac{Q'_{1}}{Q_{1}}\left(\frac{{\dot{S}}_{1}}{S_{1}} - \frac{{\dot{S}}_{4}}{S_{4}}\right) + \frac{Q'_{2}}{Q_{2}}\left(\frac{{\dot{S}}_{2}}{S_{2}} - \frac{{\dot{S}}_{4}}{S_{4}}\right) + \frac{Q'_{3}}{Q_{3}}\left(\frac{{\dot{S}}_{3}}{S_{3}} - \frac{{\dot{S}}_{4}}{S_{4}}\right) = 0,
\end{equation}

\bigskip

\begin{equation}
\label{R11}
\epsilon S_{4}^2\left[\frac{{\ddot{S}}_{1}}{S_{1}} + \frac{{\dot{S}}_{1}}{S_{1}}\left(\frac{\dot{S}_{2}}{S_{2}} + \frac{\dot{S}_{3}}{S_{3}} + \frac{\dot{S}_{4}}{S_{4}}\right)\right] + Q_{0}^2\left[\frac{{{Q}}_{1}''}{Q_{1}} + \frac{{{Q}}_{1}'}{Q_{1}}\left(\frac{{Q}_{0}'}{Q_{0}} + \frac{{Q}_{2}'}{Q_{2}} + \frac{{Q}_{3}'}{Q_{3}}\right)\right] = 0, 
\end{equation}

\bigskip

\begin{equation}
\label{R22}
\epsilon S_{4}^2\left[\frac{{\ddot{S}}_{2}}{S_{2}} + \frac{{\dot{S}}_{2}}{S_{2}}\left(\frac{\dot{S}_{1}}{S_{1}} + \frac{\dot{S}_{3}}{S_{3}} + \frac{\dot{S}_{4}}{S_{4}}\right)\right] + Q_{0}^2\left[\frac{{{Q}}_{2}''}{Q_{2}} + \frac{{{Q}}_{2}'}{Q_{2}}\left(\frac{{Q}_{0}'}{Q_{0}} + \frac{{Q}_{1}'}{Q_{1}} + \frac{{Q}_{3}'}{Q_{3}}\right)\right] = 0, 
\end{equation}

\bigskip

\begin{equation}
\label{R33}
\epsilon S_{4}^2\left[\frac{{\ddot{S}}_{3}}{S_{3}} + \frac{{\dot{S}}_{3}}{S_{3}}\left(\frac{\dot{S}_{1}}{S_{1}} + \frac{\dot{S}_{2}}{S_{2}} + \frac{\dot{S}_{4}}{S_{4}}\right)\right] + Q_{0}^2\left[\frac{{{Q}}_{3}''}{Q_{3}} + \frac{{{Q}}_{3}'}{Q_{3}}\left(\frac{{Q}_{0}'}{Q_{0}} + \frac{{Q}_{1}'}{Q_{1}} + \frac{{Q}_{2}'}{Q_{2}}\right)\right] = 0. 
\end{equation}
In these, a dot and prime mean the total derivative with respect to $t$ and $\psi$, respectively. It should be noted that the last three equations are invariant under the interchanges $1 \leftrightarrow 2$, $1 \leftrightarrow 3$, $2 \leftrightarrow 3$, as expected.

We now proceed to construct the most general solution to equations (\ref{R00})-(\ref{R33}). Towards this end, it is helpful to notice a property shared by all of them except the cross-component $R_{04}$ given by (\ref{R04}). Namely, that the functions in square brackets which multiply $S_{4}^2$ and $Q_{0}^2$ must be proportional to each other, since otherwise we obtain the trivial result that $S_{4} = 0$ and $Q_{0} = 0$. Thus, from 
(\ref{R11}) and (\ref{R22}) we find
\begin{equation}
\label{proportionality between metric functions}
\left[\frac{{\ddot{S}}_{1}}{S_{1}} + \frac{{\dot{S}}_{1}}{S_{1}}\left(\frac{\dot{S}_{2}}{S_{2}} + \frac{\dot{S}_{3}}{S_{3}} + \frac{\dot{S}_{4}}{S_{4}}\right)\right] = k \left[\frac{{\ddot{S}}_{2}}{S_{2}} + \frac{{\dot{S}}_{2}}{S_{2}}\left(\frac{\dot{S}_{1}}{S_{1}} + \frac{\dot{S}_{3}}{S_{3}} + \frac{\dot{S}_{4}}{S_{4}}\right)\right], 
\end{equation}
where $k$ is an arbitrary separation constant which is physically dimensionless (i.e., it is a pure number). Further, using the invariance under $1 \leftrightarrow 2$ mentioned before, we obtain from (\ref{proportionality between metric functions}) the relation

\begin{equation}
\label{using the symmetry}
  \left[\frac{{\ddot{S}}_{2}}{S_{2}} + \frac{{\dot{S}}_{2}}{S_{2}}\left(\frac{\dot{S}_{1}}{S_{1}} + \frac{\dot{S}_{3}}{S_{3}} + \frac{\dot{S}_{4}}{S_{4}}\right)\right] = k \left[\frac{{\ddot{S}}_{1}}{S_{1}} + \frac{{\dot{S}}_{1}}{S_{1}}\left(\frac{\dot{S}_{2}}{S_{2}} + \frac{\dot{S}_{3}}{S_{3}} + \frac{\dot{S}_{4}}{S_{4}}\right)\right].
\end{equation}
This may be combined with (\ref{proportionality between metric functions}) to give
\begin{equation}
\label{equations for S}
  \left[\frac{{\ddot{S}}_{2}}{S_{2}} + \frac{{\dot{S}}_{2}}{S_{2}}\left(\frac{\dot{S}_{1}}{S_{1}} + \frac{\dot{S}_{3}}{S_{3}} + \frac{\dot{S}_{4}}{S_{4}}\right)\right] (1 - k^2) = 0.
\end{equation}
Since $k$ is an arbitrary number, it follows that 
\begin{equation}
\label{Final equations for S}
  \left[\frac{{\ddot{S}}_{2}}{S_{2}} + \frac{{\dot{S}}_{2}}{S_{2}}\left(\frac{\dot{S}_{1}}{S_{1}} + \frac{\dot{S}_{3}}{S_{3}} + \frac{\dot{S}_{4}}{S_{4}}\right)\right] = 0.
\end{equation}
This relation has analogs that follow from the preceding argument applied to other functions  multiplying $S^2_{4}$, so there are similar functional forms for all the scale functions $S_{i}$ $(i = 1, 2, 3, 4)$. Furthermore,  (\ref{Final equations for S}) is an ordinary differential equation in $x^{0} = t$, with no appearance of $x^{4} = \psi$. This means that the solution to the field equations takes the usual Kasner form:

\begin{equation}
\label{Kasner metric in 5D, no psi dependence}
dS^2 = dt^2 - A_{1}\; t^{2\alpha_{1}}dx^2 - A_{2} \; t^{2\alpha_{2}}dy^2 - A_{3}\;  t^{2\alpha_{3}}dz^2  + \epsilon A_{4} \; t^{2\alpha_{4}}d\psi^2.
\end{equation} 
Here the  $A_{i}$ are arbitrary constants whose physical dimensions depend on those assigned to the coordinates. The  $\alpha_{i}$ are dimensionless parameters whose values are constrained by the field equations to satisfy the conditions:
\begin{equation}
\label{conditions on alpha}
\alpha_{1} + \alpha_{2} + \alpha_{3} + \alpha_{4} = 1, \;\;\;\;\alpha_{1}^2 + \alpha_{2}^2 + \alpha_{3}^2 + \alpha_{4}^2 = 1.
\end{equation}
Relations like this and its associated metric (\ref{Kasner metric in 5D, no psi dependence}) also follow when we apply the preceding analytic argument not to $S_{4}^2$ but to $Q_{0}^2$. The analog of the $t$-dependent solution (\ref{Kasner metric in 5D, no psi dependence})
is the $\psi$-dependent one 
\begin{equation}
\label{Kasner metric in 5D, no t dependence}
dS^2 = B_{0}\;  \psi^{2\beta_{0}}dt^2 - B_{1} \; \psi^{2\beta_{1}}dx^2 - B_{2} \; \psi^{2\beta_{2}}dy^2 - B_{3}\;  \psi^{2\beta_{3}}dz^2  + \epsilon\;  d\psi^2,
\end{equation}
and the analogs of the conditions (\ref{conditions on alpha}) are

\begin{equation}
\label{conditions on beta}
\beta_{0} + \beta_{1} + \beta_{2} + \beta_{3} = 1, \;\;\;\;\beta_{0}^2 + \beta_{1}^2 + \beta_{2}^2 + \beta_{3}^2 = 1.
\end{equation}
At this point we have obtained two special Kasner-type solutions, namely the $t$-form (\ref{Kasner metric in 5D, no psi dependence}) and the $\psi$-form (\ref{Kasner metric in 5D, no t dependence}).  It remains to combine these to obtain the general solution which we seek. 

\medskip

Such a general solution of the field equations (\ref{R00})-(\ref{R33}) may be verified to be given by the following form:

\begin{equation}
\label{Kasner-like metric in 5D, t and psi dependence}
dS^2 = A\;  \psi^{2\beta_{0}}dt^2 - B\; \psi^{2\beta_{1}}\; t^{2\alpha_{1}}dx^2 - C \; \psi^{2\beta_{2}}\; t^{2\alpha_{2}}dy^2 - D\;  \psi^{2\beta_{3}}\; t^{2 \alpha_{3}}dz^2  + \epsilon E\; t^{2 \alpha_{4}}  d\psi^2.
\end{equation} 
Here the constants $A - E$ are arbitrary. The parameters $\alpha_{1, 2, 3, 4}$ and $\beta_{0, 1, 2, 3}$ satisfy (\ref{conditions on alpha}) and 
(\ref{conditions on beta}) as before; but in addition the cross component $R_{04} = 0$ of (\ref{R04}) now imposes an additional condition:

\begin{equation}
\label{conditions on alpha and beta}
\alpha_{1} \beta_{1} + \alpha_{2}\beta_{2} + \alpha_{3} \beta_{3} - \beta_{0}\left(\alpha_{1} + \alpha_{2} + \alpha_{3}\right) - \alpha_{4}\left(\beta_{1} + \beta_{2} + \beta_{3}\right) = 0.
\end{equation}
This condition, together with (\ref{conditions on alpha}) and 
(\ref{conditions on beta}), might repay a detailed investigation, in order to explore the parameter space that can be occupied by the dimensionless  constants which define the general solution (\ref{Kasner-like metric in 5D, t and psi dependence}). The latter extends our understanding of Bianchi type-I solutions of Kasner type, in the sense of adding to what is known in $4D$ \cite{Kasner}, \cite{mixmaster}, \cite{limits} and generalizing what is known in $5D$ \cite{other}-\cite{ones}. This is particularly true insofar as (\ref{Kasner-like metric in 5D, t and psi dependence}) `overlaps' in form the metric for the standard $5D$ cosmologies \cite{standard}. We proceed  to make some comments on this and related issues.

The class of solutions (\ref{Kasner-like metric in 5D, t and psi dependence}) is Ricci-flat $(R_{AB} = 0)$, but in general it is not Riemann-flat (i.e., $R_{ABCD} \neq  0$). The distinction is important because certain solutions of $5D$ relativity with high degrees of symmetry may have $R_{ABCD} = 0$ and be flat in $5D$, while possessing  curved subspaces in $4D$. This is the case for the standard $5D$ cosmologies \cite{standard}, to which we will return shortly. 

In general we note that for our solution  (\ref{Kasner-like metric in 5D, t and psi dependence}) 
a  typical component of the Riemann tensor in $5D$ is proportional to $\left(M t^{2} \psi^{2\beta_{0}} + N t^{2 \alpha_{4}}\psi^2\right)$, where $M$ and $N$ are some combinations of the parameters $\alpha$ and $\beta$.  Thus, in general $R_{ABCD} \neq 0$. However, for $ \alpha_{4} = \beta_{0} = 1$, one can choose the parameters in such a way as to make $R_{ABCD} = 0$.

Clearly, from (\ref{conditions on alpha}) and (\ref{conditions on beta}) the conditions $\alpha_{4} = \beta_{0} = 1$ imply that $\alpha_{1} = \alpha_{2} = \alpha_{3} = \beta_{1} = \beta_{2} = \beta_{3} = 0$ (assuming that the metric functions are real). However, going back to the field equations we find that $R_{1}^{1} = R_{2}^{2} = R_{3}^{3} = 0 $ requires isotropic expansion for the $3D$ sections of the metric, which entails 
\begin{equation}
\label{Riemann flat model}
\alpha_{1} = \alpha_{2} = \alpha_{3} \equiv \frac{1}{a};  \;\;\;\; \beta_{1} = \beta_{2} = \beta_{3} \equiv \beta, \;\;\;\;\epsilon\left(\frac{E}{a^2}\right) + A \;\beta^2 = 0.
\end{equation}
These values with the $R_{04}$ component of the field equation (\ref{R04}) imply  $\beta = 1/(1 - a)$, where the constant $a$ is defined in (\ref{Riemann flat model}), from which we also get $E = - \epsilon A a^2/(1 - a)^2$. Drawing together these results allows us to rewrite the metric (\ref{Kasner-like metric in 5D, t and psi dependence}) as 
\begin{equation}
\label{Riemann flat 5D solution}
dS^2 = A\psi^2 dt^2 - B \psi^{2/(1 - a)}t^{2/a}\left(dx^2 + dy^2 + dz^2\right) - \frac{A a^2 t^2}{(1 - a)^2}d\psi^2.
\end{equation}
It is important to  note that  $a$ is a free parameter, i.e., it is not constrained by the field equations. The constants $A$ and $B$ can in principle be absorbed  by a suitable choice of units, when the metric becomes identical to that for the standard $5D$ cosmologies with flat space sections \cite{standard}. These models have been much discussed (see \cite{particle physics}, p. $35$ for a review). On the hypersurfaces $x^{4} = \psi = $ constant, they give back the conventional FRW models for the early radiation-like universe $(a = 2)$ and the late dust-like universe $(a = 3/2)$. These properties of matter, and other aspects of the solution (\ref{Riemann flat 5D solution}), are very special in the present context, however, because that case is the only member of the general class (\ref{Kasner-like metric in 5D, t and psi dependence}) which is Riemann-flat. 

The properties of matter for the general metric (\ref{Kasner-like metric in 5D, t and psi dependence}) may be evaluated using the standard technique based on Campbell's theorem \cite{Campbell}, \cite{Seahra}. In the present situation this consists  in isolating the $4D$ part of the relevant $5D$ geometric quantities, using them to construct the $4D$ Einstein tensor $G_{\alpha\beta}$  $(\alpha, \beta = 0, 1, 2, 3)$ and using the field equations of general relativity $G_{\alpha \beta} = 8 \pi T_{\alpha\beta }$ to identify the effective $4D$ energy-momentum tensor (see section $1$). The working is tedious, so we just quote the results:

\begin{eqnarray}
\label{EMT for homogeneous solution}
8\pi T_{0}^{0} &=& \frac{\alpha_{1}(\alpha_{2} + \alpha_{3}) + \alpha_{2}\alpha_{3}}{A \psi^{2\beta_{0}}t^2}, \nonumber \\
8\pi T_{1}^{1} &=& \frac{\alpha_{2}^2 + \alpha_{3}^2 + \alpha_{2}(\alpha_{3} - 1) - \alpha_{3}}{A \psi^{2\beta_{0}}t^2}, \nonumber \\
8\pi T_{2}^{2} &=& \frac{\alpha_{1}^2 + \alpha_{3}^2 + \alpha_{1}(\alpha_{3} - 1) - \alpha_{3}}{A \psi^{2\beta_{0}}t^2}, \nonumber \\
8\pi T_{3}^{3} &=& \frac{\alpha_{1}^2 + \alpha_{2}^2 + \alpha_{1}(\alpha_{2} - 1) - \alpha_{2}}{A \psi^{2\beta_{0}}t^2}.
\end{eqnarray}
Here the parameters are constrained by (\ref{conditions on alpha}), (\ref{conditions on beta}) and (\ref{conditions on alpha and beta}). We note that the last three components are invariant under the interchanges $1 \leftrightarrow 2$, $1 \leftrightarrow 3$, $2 \leftrightarrow 3$, a property that they inherit from the field equations (\ref{R11}), (\ref{R22}) and (\ref{R33}). Another property of (\ref{EMT for homogeneous solution}) is that, for a general choice of parameters, the components satisfy $T = T_{0}^{0} + T_{1}^{1} + T_{2}^{2} +  T_{3}^{3} = 0$. This in $4D$ physics is usually taken to mean that the equation of state is that of radiation (i.e., photons with zero rest mass) or ultra-relativistic matter (i.e., particles with finite rest mass moving close to the speed of light). However, for the so-called induced-matter theory, it is well-known that when the $4D$ part of the $5D$ metric is independent of $x^{4}$, the effective energy-momentum tensor has 
$T = 0$ (see e.g. \cite{particle physics}). Our result implies that the reverse implication need not hold.  We also note that the isotropic cosmology (\ref{Riemann flat model}) has 
\begin{equation}
\label{ETM for the isotropic case}
T = 2(2 - a)\; T_{0}^0, \;\;\; 8\pi T_{0}^{0} = \frac{3}{a^2 A\psi^2 t^2}, \;\;\; T_{1}^{1} = T_{2}^{2} = T_{3}^{3} = \left(1 - \frac{2 a}{3}\right)\; T_{0}^{0}.
\end{equation}
Since $a$ is an arbitrary parameter we have $T \neq 0$, except for radiation ($a = 2$), as expected. 

From (\ref{EMT for homogeneous solution}) it follows that $T^{1}_{1} = T^{2}_{2} = T^{3}_{3}$ requires either $\alpha_{1} = \alpha_{2} = \alpha_{3}$ or $\alpha_{1} + \alpha_{2} + \alpha_{3} = 1$. But the former choice leads to the isotropic solution (\ref{Riemann flat 5D solution}), while the latter choice implies $\alpha_{4} = 0$ by (\ref{conditions on alpha}) and leads back  to the $4D$ vacuum Kasner solution \cite{Kasner}, \cite{mixmaster}. This means that the $4D$ effective matter source for the $5D$ solution (\ref{Kasner-like metric in 5D, t and psi dependence}) cannot in general be modeled as a perfect fluid. This is consistent with other comments on the comparable $4D$ solutions. For example, one authoritative text refers to the related mixmaster cosmology as showing ``properties of empty space reminiscent of an elastic solid" (see \cite{mixmaster} p. 806). 

In addition to providing 5D analogs of the 4D Kasner solution, our metrics may also be related to others in the braneworld scenario \cite{particle physics}, \cite{algebraically}, \cite{strength}, \cite{review}. 
 In that approach there is a singular hypersurface that defines spacetime, and the properties of matter in that hypersurface may not be  
identical to the effective ones we have calculated above from the induced-matter approach as based on Campbell's theorem \cite{Campbell},  \cite{Seahra}.  
Our solutions may be verified by the application of a program such as GR-Tensor to satisfy Einstein's equations. 
 However, it is well known that in higher-dimensional physics a 5D metric may under coordinate transformations correspond to  different 4D sources. 
For this and other reasons, we recommend further investigation of the matter and dynamics associated with our general class (\ref{Kasner-like metric in 5D, t and psi dependence}) of solutions.

\section{Summary}

We have derived exact solutions of the five-dimensional field equations which extend those of general relativity that are of Bianchi type-I. Our general solution (\ref{Kasner-like metric in 5D, t and psi dependence}) has a number of parameters that are constrained by relations (\ref{conditions on alpha}), (\ref{conditions on beta}) and (\ref{conditions on alpha and beta}). It has similarities with the Kasner metric of Einstein's theory, and can be regarded as a higher-dimensional version of that model. 

It is logical to apply our solutions to the era after the big-bang, because the solutions (\ref{Kasner-like metric in 5D, t and psi dependence}) have a $4D$ source (\ref{EMT for homogeneous solution}) with a radiation-like equation of state, and because  the special form (\ref{Riemann flat 5D solution})  is known to provide a reasonable description of the universe at later epochs \cite{particle physics}, \cite{standard}. We could, for example, imagine a model which proceeds through a series of physical phase changes, the phases corresponding to different values for the assignable parameters in the metric (\ref{Kasner-like metric in 5D, t and psi dependence}). One such model has already been investigated  \cite{astrophysical}, in order to explain the transition from deceleration to acceleration suggested by recent supernova data \cite{Perlmutter}. Other consequences of $5D$ models have been studied \cite{Wesson}, in order to avoid some of the problems to do with the formation of galaxies and other structures in conventional $4D$ cosmology \cite{implications}. We have not, for want of space, investigated these issues here, but they are worthy of attention. Likewise, it has been suggested that particles might be produced in the early stages of an anisotropic universe \cite{mixmaster}, and the scalar field represented by the last component of a $5D$ metric would provide a natural mechanism for matter production.

\end{document}